\documentclass[preprint,showpacs,aps,floatfix]{revtex4} 
\usepackage{epic}
\usepackage{graphpap}
\usepackage{eepic}
\usepackage{color}
\usepackage{graphicx}
 
\begin{document}  

\title{Evaluation of the Multiplane Method for Efficient \\
Simulations of Reaction Networks} 
\author{Baruch Barzel$^1$, Ofer Biham$^1$ and Raz Kupferman$^2$}  
\affiliation{  
$^1$Racah Institute of Physics,   
The Hebrew University,   
Jerusalem 91904,   
Israel  
\\
$^2$Institute of Mathematics,   
The Hebrew University,   
Jerusalem 91904,   
Israel}

\newcommand{\N}[1]
{
\langle N_{#1} \rangle
}

\newcommand{\Nij}[2]
{
\langle N_{#1}N_{#2} \rangle
}

\newcommand{\Nijk}[3]
{
\langle N_{#1}N_{#2}N_{#3} \rangle
}

\newcommand{\Ns}[1]
{
\langle N_{#1}^2 \rangle
}

\newcommand{\TreeNetwork}[1][0.5]
{ 
\setlength{\unitlength}{#1cm}
\begin{picture}(9,8)(0.5,0.5)
\allinethickness{0.5mm}
\put(2,2){\circle{1.5}}
\put(1.25,1.25){\makebox(1.5,1.5){$X_2$}}
\put(5,7){\circle{1.5}}
\put(4.25,6.25){\makebox(1.5,1.5){$X_1$}}
\put(8,2){\circle{1.5}}
\put(7.25,1.25){\makebox(1.5,1.5){$X_3$}}
\put(2.386,2.643){\line(3,5){2.2282}}
\put(2,4){\makebox(1.5,1.5){$X_4$}}
\put(5.386,6.357){\line(3,-5){2.2282}}
\put(6.5,4){\makebox(1.5,1.5){$X_5$}}
\put(1,8){\makebox(1,1)[c]{\Large (a)}}
\end{picture}
}

\newcommand{\CorrelatedNetwork}[1][0.5]
{ 
\setlength{\unitlength}{#1cm}
\begin{picture}(9,8)(0.5,0.5)
\allinethickness{0.5mm}
\put(2,2){\circle{1.5}}
\put(1.25,1.25){\makebox(1.5,1.5){$X_2$}}
\put(5,7){\circle{1.5}}
\put(4.25,6.25){\makebox(1.5,1.5){$X_1$}}
\put(8,2){\circle{1.5}}
\put(7.25,1.25){\makebox(1.5,1.5){$X_3$}}
\put(2.386,2.643){\line(3,5){2.2282}}
\put(2,4){\makebox(1.5,1.5){$X_3$}}
\put(5.386,6.357){\line(3,-5){2.2282}}
\put(6.5,4){\makebox(1.5,1.5){$X_5$}}
\put(1,8){\makebox(1,1)[c]{\Large (b)}}
\end{picture}
}

\begin{abstract}  

Reaction networks in the bulk and on surfaces are widespread 
in physical, chemical and biological systems. 
In macroscopic systems, which include large populations of
reactive species, stochastic fluctuations are negligible 
and the reaction rates can be evaluated using rate equations.
However, many physical systems
are partitioned into microscopic domains, 
where the number of molecules in each
domain is small and fluctuations are strong.
Under these conditions, the simulation of reaction networks 
requires stochastic methods
such as direct integration of 
the master equation.
However, direct integration of
the master equation 
is infeasible for complex networks,
because the number of equations proliferates as the  
number of reactive species increases. 
Recently, the multiplane method,
which provides a dramatic reduction in the number of equations,
was introduced   
[A. Lipshtat and O. Biham, Phys. Rev. Lett. 93, 170601 (2004)].
The reduction is achieved by breaking the network into a set 
of maximal fully connected sub-networks (maximal cliques).
Lower-dimensional master equations are constructed for the
marginal probability distributions associated with the cliques,
with suitable couplings between them.
In this paper we test the multiplane method and examine its
applicability.
We show that the method is accurate in the limit
of small domains, where fluctuations are strong.
It thus provides an efficient framework for
the stochastic simulation 
of complex reaction networks with strong fluctuations,
for which rate equations fail
and direct integration of the master equation is infeasible.
The method also applies in the case of 
large domains, where it converges to
the rate equation results.

\end{abstract} 
  
 
\maketitle  

\section{Introduction}
\label{sec:introduction}

Reaction networks commonly appear
in physical, chemical and biological systems,
where reactions may take place in the bulk or on a surface.
When the surface or bulk systems are macroscopic, 
the populations of reactive species are typically large and
the law of large numbers applies.
Thus, fluctuations in the concentrations
of the reactants
and in the reaction rates become negligible. 
As a result, these reaction networks can be analyzed using rate
equation models, which account for the average concentrations
and ignore fluctuations.

In some cases, the system is partitioned 
into small domains, such that 
the reactants cannot diffuse between them.
The populations of reactive species in
each domain become small and their fluctuations cannot be ignored.
As a consequence, rate equations fail 
and the simulation of these reactions
requires stochastic methods such as direct integration 
of the master equation
\cite{vanKampen1981}.
The master equation takes into account the discrete nature of
the reactants as well as the fluctuations.
It is expressed in terms of the probabilities of having a given
set of population sizes of the reactive species
in a given domain. 
In certain cases, such as radioactive decay, an analytical solution
based on generating functions is available
\cite{McQuarrie1967}.
In other cases, numerical methods are required.
For simple reaction networks that involve few reactive species,
numerical integration of the master equation 
is useful and efficient
\cite{Biham2001,Green2001,Biham2002}.
However, as the number of reactive species increases,
the number of variables in the master equation
quickly proliferates
\cite{Stantcheva2002,Stantcheva2003},
making the direct integraion 
infeasible.

Here we focus on networks in which reactions take place between pairs
of species, and the reaction products may be reactive or non-reactive.
Such networks may be described by graphs: 
each reactive species is represented by a node; 
the reaction between a pair of species 
is represented by an edge that connects
the corresponding nodes.
Typically, these networks are sparse, namely most 
pairs of species do not react with each other.
For such sparse networks, the recently introduced multiplane method 
provides a dramatic reduction in the number of equations
\cite{Lipshtat2004}.
The method is based on breaking the network into a set of 
maximal fully connected subnetworks (maximal cliques).
It involves an approximation, 
in which the correlations between 
pairs of species that react with each other are maintained, while
the correlations between non-reacting pairs are neglected. 
The result is a set of lower dimensional master equations,
one for each clique, with suitable couplings between them.
For sparse networks, the cliques are typically small and mostly
consist of two or three nodes.
This method thus enables the simulation of large networks much
beyond the point where the master equation becomes infeasible.

The multiplane method has already been used in the simulation of 
complex chemical networks on dust grains in interstellar clouds
\cite{Lipshtat2004},
where rate equations fail
\cite{Tielens1982,Charnley1997,Caselli1998,Shalabiea1998},
while
direct integration
of the master equation is impractical
\cite{Stantcheva2002,Stantcheva2003}.
The multiplane method is also required
for the simulation of 
genetic networks in cells,
where
the master equation
\cite{Paulsson2000,Paulsson2000b,Paulsson2004} 
and Monte Carlo simulations
\cite{Gillespie1977,Mcadams1997,Mcadams1999}
are not applicable for large networks.

In this paper we analyze the multiplane method
and examine its validity.
This is done by comparing the results with the complete
master equation.
The comparison is done both for the probability distributions 
and for the first and second moments, which represent the
population sizes of reactants and reaction rates, respectively. 
It is shown that the multiplane method provides accurate results
for both the population sizes and reaction rates.
For concreteness, we use below the terminology of surface reactions.
In this context, 
the small domains are taken to be dust grains
(assumed to be spherical, for simplicity),
and the reactants are atoms or molecules
that enter the system as incoming flux from the surrounding gas phase
(below we use the words atoms and molecules interchangeably).
The reactants and reaction products
leave the system by thermal desorption.
The reactions that take place on a grain are driven by diffusion
of reactants on its surface until they encounter each other and react.
In spite of this specific terminology, the multiplane method can 
be adapted to other contexts, such as reactions in a solution,
protein interactions in a living cell and birth-death processes
in population dynamics. 
Here we focus on the calculation of steady-state
solutions and thus do not expand on birth-death processes which
may exhibit absorbing states.

The paper is organized as follows. 
In Sec.
\ref{rate_master}
we briefly review the rate equation and master equation methods,
presenting them for a simple reaction network.
In Sec.
\ref{the_method}
we describe the multiplane method.
In Sec. \ref{simulations}
we test the performance of the multiplane method.
An analysis of the method in the limits of 
small and large grains is presented in
Sec. \ref{analysis}.
In Sec.
\ref{complex}
we show how to apply the method to more complex networks.
In Sec.
\ref{applications}
we briefly describe 
its applications in interstellar chemistry and in genetic
networks.
The main findings are summarized and discussed in 
Sec. \ref{summary}.

\section{The Rate Equations and the Master Equation}
\label{rate_master}

Consider a small grain,
exposed to fluxes of three different atomic species,
denoted by
$X_1$,
$X_2$
and
$X_3$.
The adsorbed atoms on the grain reside in adsorption sites.
The number of sites, $S$, is proportional to the surface
area of the grain.
The incoming flux of 
$X_i$, $i = 1, 2, 3$,
is given by
$f_i$ (s$^{-1}$)
atoms per site. 
Thus, the flux of atoms per grain is 
$F_i = f_i S$ (s$^{-1}$). 
The adsorbed atoms may desorb due to thermal excitations. 
The desorption rate of the 
$X_i$ species
from the grain is denoted by
$W_i$ (s$^{-1}$).
While residing on the grain, the atoms diffuse on the
surface via hopping between adjacent sites.
The hopping rate of 
$X_i$
atoms 
is given by
$a_i$ (hops s$^{-1}$).
It is convenient to define
the sweeping rate
$A_i = a_i / S$,
which is approximately the inverse of the time it takes
an adsorbed
$X_i$
atom to visit nearly all
the adsorption sites on the grain surface
\cite{Lipshtat2002}.
A more accurate expression for $A_i$ in the case of
spherical grains appears in Ref.
\cite{Lohmar2006},
where it is shown to be reduced by a logarithmic factor.

The diffusion process enables adsorbed atoms to encounter
each other and react.
Here we consider a simple reaction network that includes
the reactions
$X_1 + X_2 \longrightarrow X_4$
and  
$X_1 + X_3 \longrightarrow X_5$,
where the $X_4$ and $X_5$ molecules are the reaction products.
The graph  
that illustrates
this network is shown in Fig. 
\ref{fig1}(a).

\subsection{The Rate Equations}

The rate equations that describe the network 
of 
Fig. \ref{fig1}(a).
take the form

\begin{eqnarray}
\frac{d \N{1}}{dt} &=&
F_1 - W_1\N{1} - (A_1 + A_2)\N{1}\N{2} - (A_1 + A_3)\N{1}\N{3}
\nonumber \\
\frac{d \N{2}}{dt} &=&
F_2 - W_2\N{2} - (A_1 + A_2)\N{1}\N{2} 
\nonumber \\
\frac{d \N{3}}{dt} &=&
F_3 - W_3\N{3} - (A_1 + A_3)\N{1}\N{3}, 
\label{eq:rate3species}
\end{eqnarray} 

\noindent
where $\N{i}$ is the
average population size of 
$X_i$ 
atoms on a grain.
The first terms on the right hand side of Eq.
(\ref{eq:rate3species}) 
represent the incoming
fluxes of 
$X_i$ atoms.
The second terms 
represent the desorption of 
$X_i$ 
atoms, which is proportional to the  
$X_i$ 
population on the grain.
The remaining terms account for the reactions between 
adsorbed atoms.
The production rates of
$X_4$ and $X_5$
molecules per grain
(in units of s$^{-1}$)
are given by
$R_4 = (A_1 + A_2) \N{1}\N{2}$
and 
$R_5 = (A_1 + A_3) \N{1}\N{3}$.
For simplicity, we assume that non-reactive product species
desorb into the gas phase immediately upon formation.

For large grains, 
Eqs. (\ref{eq:rate3species})
account correctly for the reaction rates.
However, in the limit of small grains,
some of the average population sizes,
$\N{i}$,
may become small.
In this case the discrete nature of the adsorbed atoms
and molecules becomes important and the fluctuations
cannot be ignored.
As a result, 
the reaction rates obtained from the rate equations
(\ref{eq:rate3species}) 
are incorrect
and stochastic methods are needed.

We also consider a related network,
shown in Fig. 
\ref{fig1}(b),
in which $X_3$ is the product of the reaction between
$X_1$ and $X_2$
(namely, $X_3$ and $X_4$ are the same species).
The rate equations that describe this system
are the same as in 
Eq. (\ref{eq:rate3species})
except that in the third equation
one needs to add the term
$(A_1 + A_3)\N{1}\N{3}$.
The production rate of $X_5$ is still given by
$R_5$ defined above. 

\subsection{The Master Equation}

The dynamical variables
of the master equation are the probabilities
$P(n_1,n_2,n_3)$
of having a population of 
$n_i$ 
atoms of species 
$X_i$
on the grain. 
It takes the form

\begin{eqnarray}
\dot P(n_1,n_2,n_3) 
&=& \sum_{i=1}^{3}{F_i[P(\dots,n_i-1,\dots) - 
P(n_1,n_2,n_3)]} \nonumber \\
&+& \sum_{i=1}^{3}{W_i[(n_i + 1)P(\dots,n_i+1,\dots) - 
n_iP(n_1,n_2,n_3)]}  \\
&+& (A_1 + A_2)[(n_1 + 1)(n_2 + 1)P(n_1 + 1,n_2 + 1,n_3) - 
n_1n_2P(n_1,n_2,n_3)] \nonumber \\
&+& (A_1 + A_3)[(n_1 + 1)(n_3 + 1)P(n_1 + 1,n_2,n_3 + 1) - 
n_1n_3P(n_1,n_2,n_3)], \nonumber
\label{eq:master3species}
\end{eqnarray}

\noindent
where
$n_i = 0,1,2,\dots$.
The first term in 
Eq. (\ref{eq:master3species})
describes the effect of the incoming flux.
The probability 
$P(\dots,n_i,\dots)$ 
increases when an 
$X_i$ atom is adsorbed on
a grain that already has 
$n_i-1$ 
adsorbed
$X_i$
atoms. 
This probability decreases when an 
$X_i$ atom is adsorbed on
a grain that includes 
$n_i$ 
atoms of species
$X_i$.
The second term accounts for the desorption process. 
The third and fourth terms describe the reactions 
that take place on the grain.

In numerical simulations the master equation is 
truncated in order
to keep the number of equations finite.
A convenient way to achieve this is to assign upper cutoffs
on the population sizes of the reactive species,
$n_i^{\rm max}$, $i=1,\dots,J$, 
where $J$ is the number 
of reactive species.

In the network of 
Fig. \ref{fig1}(a),
the average population size
of $X_i$ on a grain is given by the first 
moment

\begin{equation}
\langle N_i \rangle =
\sum_{n_1=0}^{n_1^{\rm max}}
\sum_{n_2=0}^{n_3^{\rm max}}
\sum_{n_3=0}^{n_3^{\rm max}}
n_i P(n_1,n_2,n_3).
\end{equation}

\noindent
The production rates per grain of 
$X_4$
and
$X_5$ 
molecules can be obtained from the 
mixed second moments of
$P(n_1,n_2,n_3)$, 
according to
$R_4 = (A_1 + A_2) \Nij{1}{2}$
and 
$R_5 = (A_1 + A_3) \Nij{1}{3}$,
where

\begin{equation}
\langle N_i N_j \rangle =
\sum_{n_1=0}^{n_1^{\rm max}}
\sum_{n_2=0}^{n_3^{\rm max}}
\sum_{n_3=0}^{n_3^{\rm max}}
n_i n_j P(n_1,n_2,n_3).
\end{equation}

In a network of $J$ reactive species,
the number of equations to be solved is  
 
\begin{equation}
N_E = \prod_{i=1}^J (n_i^{\max}+1).
\label{eq:numofequations}
\end{equation}

\noindent
The truncated master equation is valid if the probability
to have a population larger 
than the assigned cutoff is negligible.
Note that
$N_E$ 
grows exponentially with the number of reactive species.
This limits the applicability of the master equation
to simple networks, making it impractical in the case of
complex networks which involve many reactive species
\cite{Stantcheva2002,Stantcheva2003}.

\section{The Multiplane Method}
\label{the_method}

The recently introduced multiplane 
method provides a dramatic reduction 
in the number of equations
\cite{Lipshtat2004}.
It thus enables efficient simulations of complex reaction networks.
Below we describe the method using the network of 
Fig. \ref{fig1}(a).
Note that in this network 
the species 
$X_1$ 
participates in
both reactions.
Since the species
$X_2$ 
and 
$X_3$ 
do not react with each other,
one may assume that
for a given population size of $X_1$, 
their population sizes are almost conditionally independent.
Under this assumption, the probability distribution of the 
population sizes can be approximated by
\cite{Lipshtat2004}

\begin{equation}
P(n_1,n_2,n_3) = P(n_1)P(n_2|n_1)P(n_3|n_1),
\label{eq:conditional}
\end{equation}

\noindent
where 
$P(n_i|n_1)$ 
is the conditional probability that
there will be 
$n_i$ atoms of species 
$X_i$ given that there are 
$n_1$ atoms of species 
$X_1$ on the grain. 

In order to derive the multiplane equations,
we first insert Eq.
(\ref{eq:conditional})
into the master Eq. 
(\ref{eq:master3species}), 
and trace over the population size of
$X_3$.
Using the fact that
$\sum_{n_3}{P(n_3|n_1)} = 1$
and that
$\sum_{n_3}{\dot P(n_3|n_1)} = 0$,
one obtains 

\begin{eqnarray}
\dot P(n_1,n_2)
&=& F_1[P(n_1-1,n_2) - P(n_1,n_2)] + 
F_2[P(n_1,n_2-1) - P(n_1,n_2)] \nonumber \\
&+& W_1[(n_1 + 1)P(n_1+1,n_2) - n_1P(n_1,n_2)] \nonumber \\
&+& W_2[(n_2 + 1)P(n_1,n_2+1) - n_1P(n_1,n_2)] \nonumber \\
&+& (A_1 + A_2)[(n_1 + 1)(n_2 + 1)P(n_1 + 1,n_2 + 1) - 
n_1n_2P(n_1,n_2)]  \nonumber \\
&+& (A_1 + A_3)[(n_1 + 1)\N{3}_{n_1+1}P(n_1 + 1,n_2) - 
n_1\N{3}_{n_1}P(n_1,n_2)],
\label{eq:mp3species1}
\end{eqnarray}
 
\noindent
where 
$\N{3}_{n_1} = \sum_{n_3}{n_3P(n_3|n_1)}$.
A similar procedure,  
tracing over the population size of
$X_2$,
leads to the equation

\begin{eqnarray}
\dot P(n_1,n_3)
&=& F_1[P(n_1-1,n_3) - P(n_1,n_3)] + 
F_3[P(n_1,n_3-1) - P(n_1,n_3)] \nonumber \\
&+& W_1[(n_1 + 1)P(n_1+1,n_3) - n_1P(n_1,n_3)] \nonumber \\
&+& W_3[(n_3 + 1)P(n_1,n_3+1) - n_1P(n_1,n_3)] \nonumber \\
&+& (A_1 + A_3)
[(n_1 + 1)(n_3 + 1)P(n_1 + 1,n_3 + 1) - n_1n_3P(n_1,n_3)]  \nonumber\\
&+& (A_1 + A_2)
[(n_1 + 1)\N{2}_{n_1+1}P(n_1 + 1,n_3) - n_1\N{2}_{n_1}P(n_1,n_3)].
\label{eq:mp3species2}
\end{eqnarray}
 
\noindent
These are, in fact, two master equations, one for 
$P(n_1,n_2)$
and the other for 
$P(n_1,n_3)$.
These two master equations are coupled through the conditional averages
$\N{j}_{n_1}$
where 
$j=2,3$.
The conditional average, which is evaluated in each 
one of these master equations
is then used, essentially as a rate constant, 
in the other master equation. 
The multiplane equations are solved by direct numerical integration
using standard steppers such as Runge-Kutta.
At each time step, the probability distributions 
$P(n_1,n_2)$ 
and 
$P(n_1,n_3)$ 
are updated.
The conditional averages
$\N{j}_{n_1}$
are then evaluated and used in the next time step.

The number of equations is significantly reduced
as we replace the three-dimensional set of equations for
$P(n_1,n_2,n_3)$ 
by two-dimensional sets for
$P(n_1,n_2)$, 
and
$P(n_1,n_3)$.
The number of equations in the three-dimensional set is
given by 
Eq. (\ref{eq:numofequations})
with $J=3$.
The number of equations in each one 
of the two dimensional sets is 
$(n_1^{\max}+1) (n_i^{\max}+1)$, $i=2,3$.

The multiplane method enables one 
to calculate the average population sizes
$\N{i}$
of all the species, as well as the reaction rates, expressed in terms of
the second moments
$\Nij{i}{j}$.
Consider, for example, the population size 
$\N{1}$ 
of the 
$X_1$ 
species. It can be expressed in two ways, namely

\begin{equation}
\N{1} = \sum_{n_1 = 0}^{n_1^{\rm max}}
{\sum_{n_i = 0}^{n_i^{\rm max}}{n_1 P(n_1,n_i)}}
\end{equation}
 
\noindent
where 
$i=2$ or $3$.
In the first case 
$\N{i}$ 
is evaluated from 
$P(n_1,n_2)$,
and in the second case it is evaluated from
$P(n_1,n_3)$.
A nice property of the multiplane method is that the results are identical,
as can be seen from 
Eq. (\ref{eq:conditional}).
The difference is merely in the order in which 
$N_2$ and $N_3$
are traced out. 
The multiplane method also provides the reaction rates.
For example, the production rate of the 
$X_4$ 
species [Fig. \ref{fig1}(a)] is given by

\begin{equation}
R_4 = (A_1 + A_2) \sum_{n_1=0}^{n_1^{\rm max}} 
\sum_{n_2=0}^{n_2^{\rm max}} n_1n_2 P(n_1,n_2). 
\end{equation}

Note that in the derivation of the multiplane equations, 
certain dependencies were neglected. 
Still, the dependence between all pairs of 
species that react with each other are maintained
through the conditional averages,
$\N{i}_{n_1}$.
These conditional averages are essential 
in order to maintain the desired
correlations. 
If the conditional moments  
$\N{i}_{n_1}$,
in the multiplane equations 
(\ref{eq:mp3species1}) 
and 
(\ref{eq:mp3species2}), 
are replaced 
by 
$\N{i}$
for 
$i=2,3$, 
these equations 
are reduced, 
by proper summations, 
to the rate equations
(\ref{eq:rate3species}).
In this case all the correlations are lost.

\section{Simulations and Results}
\label{simulations}

\begin{figure}
\includegraphics[width=4.5in]{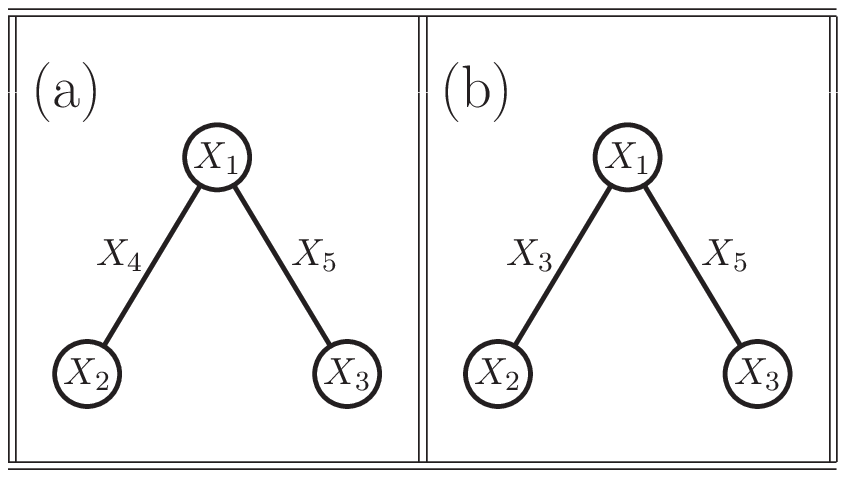}
\caption
{
Graphic representations of two reaction networks that
involve three reactive species.
The nodes represent reactive species and the edges
represent reactions between pairs of species.
The reaction products are specified near the edges.
In these networks there are two
cliques:
one consists of
$X_1$ and $X_2$
and the other consists of
$X_1$ and $X_3$.
(a) The reaction products, $X_4$ and $X_5$ 
are non-reactive;
(b) 
The product of the reaction between
$X_1$ and $X_2$
is the reactive specie
$X_3$. 
}
\label{fig1}
\end{figure}

\begin{figure} 
\includegraphics[width=5.0in]{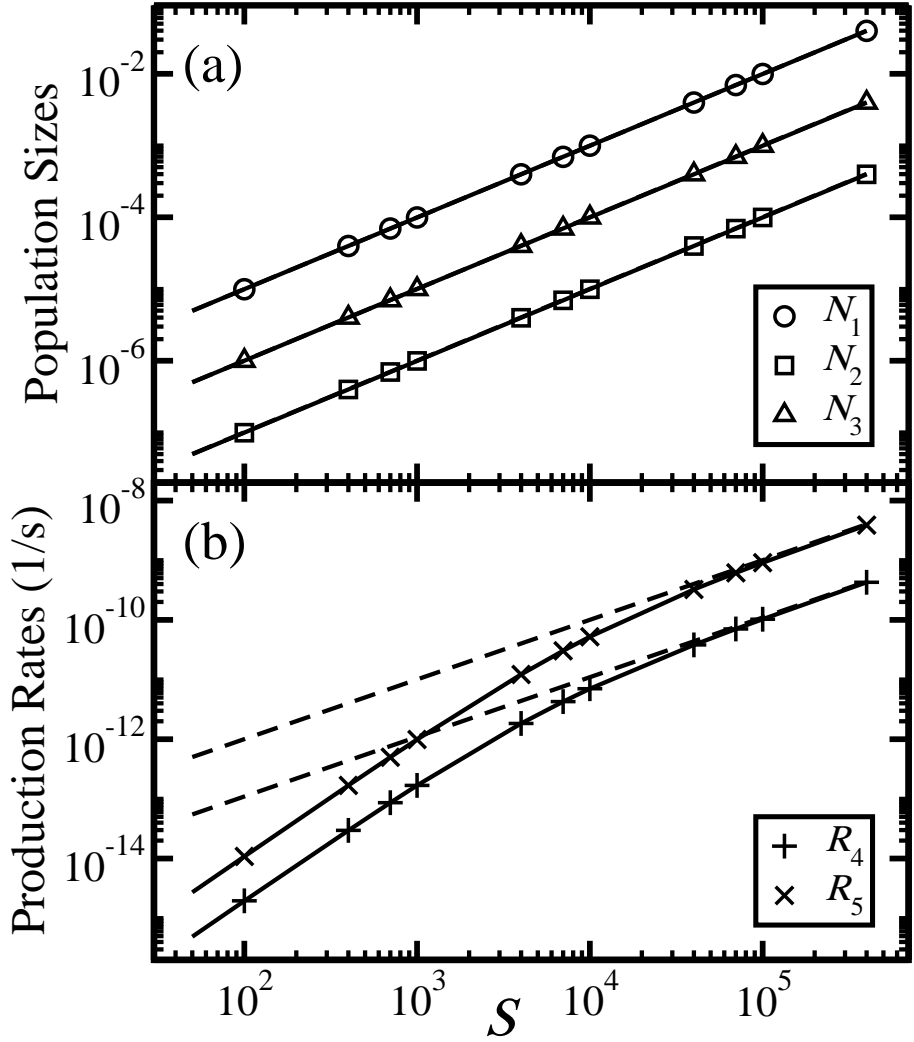}
\caption
{
(a) 
The population sizes of the 
$X_1$ (circles), 
$X_2$ (triangles)
and 
$X_3$ (squares)
species per grain vs. the number of adsorption sites, 
$S$, on the grain,
obtained from the multiplane equations,
for the network shown in 
Fig. \ref{fig1}(a).
The results are in perfect agreement with the master equation
(solid lines) and the rate equations (dashed lines);
(b) 
The production rates of 
$X_4$ (+)
and
$X_5$ ($\times$) 
molecules per grain vs. $S$,
obtained from the multiplane equations.
The results are in perfect agreement with the master equation
(solid lines).
For small grains, the rate equation results (dashed lines)
for the reaction rates exhibit large deviations.
Here,
$X_1$ is the dominant species.
}
\label{fig2}
\end{figure}

To examine the multiplane method we have performed simulations of 
the reaction networks shown in 
Fig. \ref{fig1}.
The results were compared to those obtained from the complete
master equation.
In Fig. \ref{fig2}(a) 
we present the average population sizes of the
$X_1$ (circles),
$X_2$ (squares)
and 
$X_3$ (triangles)
species on a grain vs. the number of adsorption sites, 
$S$, 
for the network of 
Fig. \ref{fig1}(a),
obtained from the multiplane equations 
under steady state conditions.
In the simulations throughout the paper,
we chose to use  
the parameters 
$W_1 = 10^{-3}$,
$a_1 = 10$,
$W_2 = 10^{-3}$,
$a_2 = 1$,
$W_3 = 10^{-5}$,
and
$a_3 = 10^{-1}$
(s$^{-1}$).
This choice reflects the mobilities and desorption rates
in the network of H, O and OH that appears in interstellar
grain chemistry
\cite{Stantcheva2001}.
The production rates of
$X_4$ ($+$)
and 
$X_5$ ($\times$)
molecules on a grain, vs. 
$S$, 
obtained from the multiplane equations,
are shown in 
Fig.
\ref{fig2}(b). 
The results are in excellent agreement with the 
master equation (solid lines).
The rate equations (dashed lines) provide accurate results
for large grains, but for small grains they show significant deviations.
We have preformed extensive simulations of this system,
using a wide range of parameters, and found that the consistency of
the multiplane method and the complete 
master equation is always maintained.
In the simulations presented above the fluxes were
$F_1 = 10^{-8} S$,
and
$F_2 = F_3 = 0.01 F_1$.

\begin{figure} 
\includegraphics[width=5.0in]{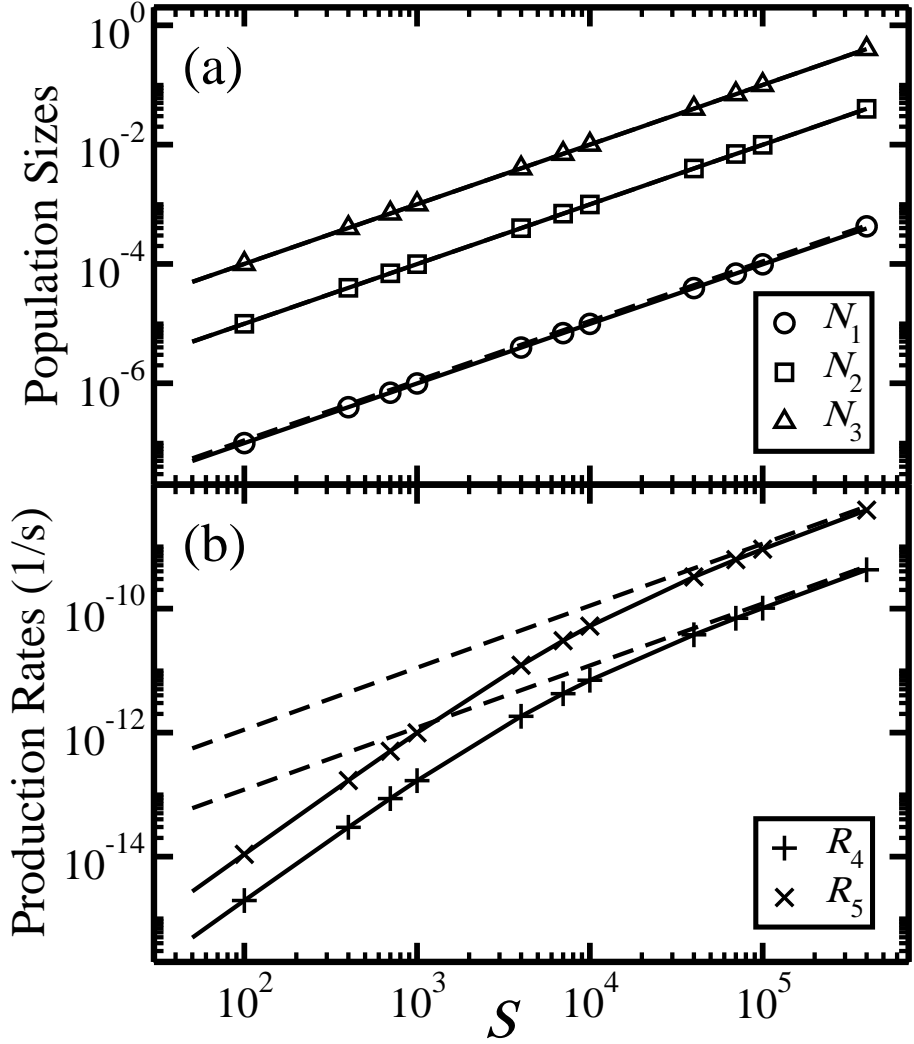}
\caption
{
The population sizes 
(a) 
and the production rates
(b)
per grain vs. $S$ 
for the network of
Fig. \ref{fig1}(a).
The multiplane results (symbols) 
are in perfect agreement with the master equation
(solid lines).
The rate equation results (dashed lines)
deviate significantly for small grains.
Here the species
$X_1$ is dominated by 
$X_2$ and
$X_3$.
}
\label{fig3}
\end{figure}

Note that with the parameters specified above, 
the incoming flux of $X_1$ atoms 
is much larger than the fluxes of $X_2$ and $X_3$.
It is often the case in chemical networks that there
exists a dominant species, which is more abundant and more reactive
than the other species 
(such as hydrogen in interstellar grain chemistry). 
One could speculate that
the dominance of $X_1$ is the reason
for the remarkable agreement between the multiplane results
and the master equation results. 
In order to show 
that this is not the case,
and that the multiplane equations are
generically applicable,
we examine some other parameters. 
In particular,
we consider the case in which the flux of $X_1$
is much lower than the fluxes of $X_2$ and $X_3$,
namely
$F_2 = F_3 = 10^{-8}$
and 
$F_1 = 0.01 F_2$. 
The population sizes and reaction rates
obtained for this choice of fluxes
are shown in 
Fig. \ref{fig3}(a) and (b), respectively. 
Clearly, the excellent agreement between the multiplane
method and the master equation is maintained
in this case as well as in all other sets of parameters
that we have examined.

It turns out that the multiplane method applies even
when one of the species in one clique is a product 
of a reaction that is included in another clique.
To demonstrate this fact we consider the
network of 
Fig. \ref{fig1}(b) 
in which
$X_3$
is the product of the reaction between  
$X_1$
and 
$X_2$.
This feature may give rise to 
some sort of correlation between the population sizes of
$X_2$
and 
$X_3$. 
The question is whether such correlations may
reduce the applicability of the multiplane method.

\begin{figure} 
\includegraphics[width=5.0in]{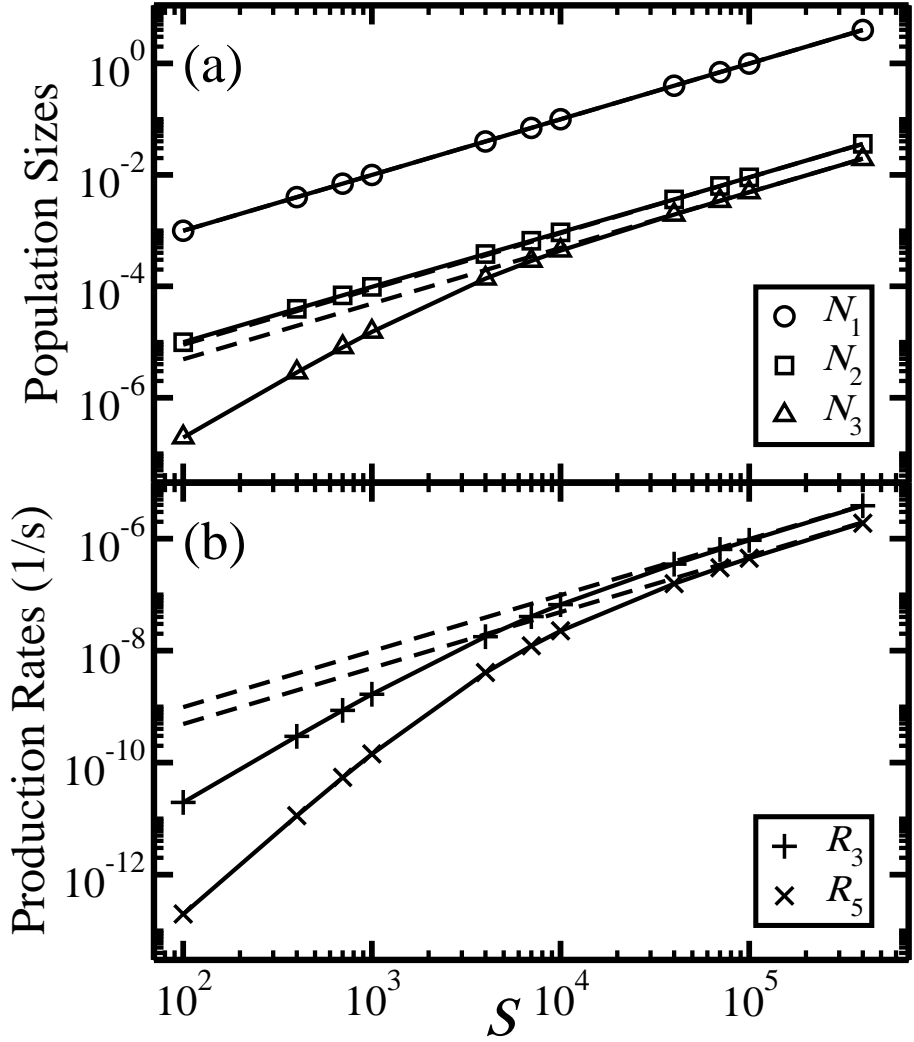}
\caption
{
The population sizes 
(a) 
and the production rates
(b)
per grain vs. $S$, 
for the network of
Fig. \ref{fig1}(b)
in which $X_3$ is the 
reaction product of $X_1$ and $X_2$.
The multiplane results (symbols) 
are in perfect agreement with the master equation
(solid lines). 
The rate equations (dashed lines) are also shown.
}
\label{fig4}
\end{figure}

The multiplane equations describing the network
of Fig. \ref{fig1}(b) 
are the same as Eqs.
(\ref{eq:mp3species1})
and
(\ref{eq:mp3species2}),
except that in the last term of the second equation,
$P(n_1+1,n_3)$
is replaced by
$P(n_1+1,n_3-1)$.
In 
Fig \ref{fig4}(a) 
we present the population sizes of the
$X_1$ (circles),
$X_2$ (triangles)
and
$X_3$ (squares)
species on a grain 
vs. $S$
under steady state conditions,
obtained from the multiplane method 
for the network of 
Fig. \ref{fig1}(b). 
The production rates of the 
$X_3$ (+)
and
$X_5$ ($\times$)
species are shown in 
Fig. \ref{fig4}(b) .
Even in this case,
the multiplane results are in perfect agreement with the 
master equation (solid lines). 
The rate equations (dashed lines) are accurate for large grains
but deviate for small grains.
Here we chose
$F_1 = 10^{-8}$,
$F_2 = 0.01 F_1$
and
$F_3 = 0$, 
namely 
$X_3$ molecules 
are not accreted from the gas phase, 
and are produced only on the grain.

Figs. 
\ref{fig2} - \ref{fig4}
demonstrate the usefulness of the multiplane method
for the simulation of reaction networks on small grains.
In particular, it is shown that the multiplane equations
provide accurate results
for the population sizes of reactants, 
given by the first moments
$\N{i}$, $i = 1,2,3$,
and for the reaction rates,
expressed in terms of the second moments,
$\Nij{1}{2}$ 
and $\Nij{1}{3}$.

The multiplane method only includes the 
marginal probability distributions,
$P(n_1,n_2)$ 
and 
$P(n_1,n_3)$.
However, using 
Eq. (\ref{eq:conditional})
one can construct an approximation of the 
complete probability distribution,
$P(n_1,n_2,n_3)$.
This approximation takes the form

\begin{equation}
P_{\rm MP}(n_1,n_2,n_3) = \frac{P(n_1,n_2) P(n_1,n_3)}
{P(n_1)},
\end{equation}

\noindent
where
the marginal probability distributions
$P(n_1,n_2)$, 
$P(n_1,n_3)$ 
and 
$P(n_1)$
on the right hand side  
are those obtained from the multiplane method.
In order to examine the accuracy of this approximation,
we introduce the {\it deviation distance},

\begin{equation}
\Delta = \sum_{n_1,n_2,n_3}{\left| P(n_1,n_2,n_3)-
P_{\rm MP}(n_1,n_2,n_3) \right|},
\label{eq:gamma}
\end{equation} 

\noindent
which is evaluated under
steady-state conditions of the master equation
and the multiplane equations,
where
$P(n_1,n_2,n_3)$
is the distribution obtained from the master equation.
We have evaluated 
$\Delta$
for a range of grain sizes between 
$S=10^2$ 
and
$S=10^6$. 
It was found 
that in all cases
$\Delta \ll 1$.
More explicitly, it varies between
$\Delta \approx 10^{-4}$ 
to 
$\Delta \approx 10^{-5}$.
This indicates that the reconstructed probability distribution
$P_{\rm MP}(n_1,n_2,n_3)$
provides a very good approximation of
$P(n_1,n_2,n_3)$.

\begin{figure} 
\includegraphics[width=4.5in]{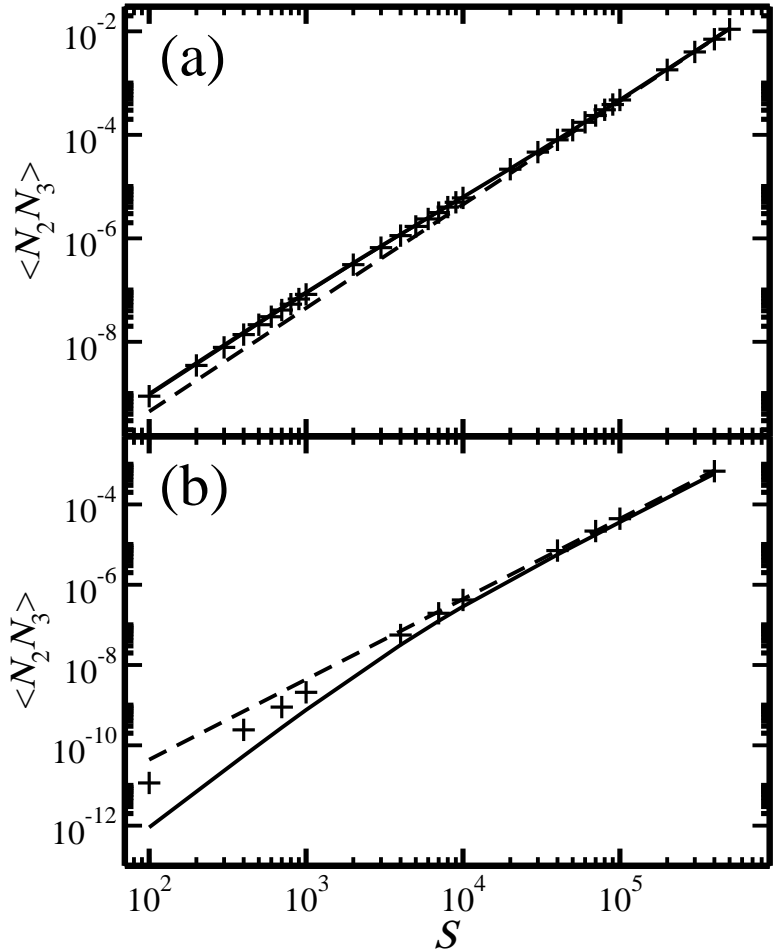}
\caption
{
(a) 
The second moment 
$\Nij{2}{3}$ 
of 
$P(n_1,n_2,n_3)$ 
vs. $S$, obtained from the multiplane method (+)
for the network shown in 
Fig. \ref{fig1}(a).
This moment is not related to any reaction rate, 
thus the multiplane method is not designed to 
approximate it well. 
Still, it turns out that
the results are in good agreement with
the master equation (solid line).
The rate equation results for
the corresponding term, 
$\N{2}\N{3}$, 
are also shown (dashed line); 
(b) 
The moment, 
$\Nij{2}{3}$, 
vs. $S$ 
for the network shown in 
Fig. \ref{fig1}(b).
The multiplane results (+) deviate from those of the
master equation (solid line)
in the limit of small grains.
For large grains the multiplane results coincide 
with the master equation and with the corresponding term,
$\N{2}\N{3}$, 
of the rate equations
(dashed line). 
}
\label{fig5}
\end{figure}

While the second moments which involve pairs of 
species in the same clique
account for their reaction rate, such moments for species
from different cliques have no direct physical interpretation.
Still, they can be used as an additional test for the accuracy of
$P_{\rm MP}(n_1,n_2,n_3)$.
Clearly, one may not expect the multiplane method to provide accurate
results for such moments because the corresponding correlations
are neglected.
In Fig. 
\ref{fig5}(a) 
we show 
the moment 
$\Nij{2}{3}$
vs. grain size as obtained from the multiplane equations
for the reaction network of 
Fig. \ref{fig1}(a) (+).
Surprisingly, the results  
are in agreement with those of the
master equation (solid line).
The corresponding rate equation results
for
$\N{2}\N{3}$,
are also shown 
(dashed line).
In Fig \ref{fig5}(b)
we show the moment 
$\Nij{2}{3}$
vs. grain size,
as obtained from the multiplane equations (+)
for the network of in 
Fig. \ref{fig1}(b).
In this network,
the species $X_3$ is the result
of the reaction between $X_1$ and $X_2$,
enhancing the correlations between them.
Indeed, 
the results of the multiplane method
deviate from the master equation results (solid line)
in the regime of small grains.
However, for large grains the results 
of the multiplane method and the master equation coincide
and agree with those of the rate equations (dashed line).
In general, we find that for higher moments
of the form
$\langle N_1^aN_2^bN_2^c \rangle$, 
$a,b,c = 1,2,\dots$,
that involve species from
more than one clique, 
the multiplane method is not reliable in the
limit of small grains. 
For large grains the multiplane and master
equation results coincide.

\section{Analysis of the Method}
\label{analysis}

\subsection{The Limit of Small Grains}

Consider the probability distribution
$P(n_1,n_2,n_3)$
in the limit of small grains, where 
the average population sizes satisfy
$\langle N_i \rangle \ll 1$ 
for $i=1,2,3$.
In this limit, 
$\langle N_i \rangle$
can be expressed by
$\langle N_i \rangle \simeq \rho_i \epsilon$,
where $\rho_i \le 1$ is a constant that 
depends on the parameters and
$\epsilon \ll 1$ is proportional to the grain size, $S$.
In this case,
$P(0,0,0)$ 
is the highest probability
while
$P(1,0,0)$,
$P(0,1,0)$
and
$P(0,0,1)$
are of order 
$\epsilon$.
The probability 
$P(0,1,1)$
of having a pair of $X_2$ and $X_3$
atoms reside simultaneously on the grain is
of order
$\epsilon^2$.
The probabilities
$P(1,1,0)$
and
$P(1,0,1)$,
of having pairs of atoms of species that react with each other 
reside simultaneously on the grain, are reduced due to the reactions
and go like
$\epsilon^3$.
Under these circumstances, 
the average population sizes satisfy

\begin{eqnarray}
\N{1} &\simeq& P(1,0,0) + \mathcal{O}(\epsilon^2) \nonumber \\
\N{2} &\simeq& P(0,1,0) + \mathcal{O}(\epsilon^2) \nonumber \\
\N{3} &\simeq& P(0,0,1) + \mathcal{O}(\epsilon^2), 
\label{eq:first_moments_small}
\end{eqnarray}

\noindent
while the second moments that determine the reaction
rates satisfy

\begin{eqnarray}
\Nij{1}{2} &\simeq& P(1,1,0) + \mathcal{O}(\epsilon^4) \nonumber \\
\Nij{1}{3} &\simeq& P(1,0,1) + \mathcal{O}(\epsilon^4). 
\label{eq:second_moments_small}
\end{eqnarray}

\noindent
Using these relations, one can
show that in the limit of small grains,
to first order in $\epsilon$,
the population sizes of $X_2$ and $X_3$
are statistically independent, namely 

\begin{equation}
P(n_2,n_3) \simeq P(n_2)P(n_3) 
+ \mathcal{O}(\epsilon^3).
\label{eq:small_conditional}
\end{equation}

\noindent
To show this relation, one needs to examine
three states of $(N_2,N_3)$,
namely
$(n_2,n_3)=(0,0)$, $(0,1)$ and $(1,0)$.
In all other cases,
$P(n_2,n_3)$ 
goes like the quadratic or a higher 
degree of $\epsilon$.
As an example, we show that 
$P(N_2=0,N_3=0) \simeq P(N_2=0)P(N_3=0)$,
to leading order in $\epsilon$.
To this end, we evaluate the left hand side

\begin{equation}
P(N_2=0,N_3=0) = P(0,0,0) + P(1,0,0)
+ \mathcal{O}(\epsilon^2),
\end{equation}

\noindent
and the right hand side

\begin{eqnarray}
P(N_2=0) P(N_3=0) &=& 
\left[ 1 - P(N_2=1) + \mathcal{O}(\epsilon^2) \right] 
\left[ 1 - P(N_3=1) + \mathcal{O}(\epsilon^2) \right] 
\nonumber \\
&=& P(0,0,0) + P(1,0,0) + \mathcal{O}(\epsilon^2).
\label{eq:expansion_small}
\end{eqnarray}
             
\noindent
Clearly, the relation
(\ref{eq:small_conditional})
is satisfied.
This result justifies the applicability of 
Eq. (\ref{eq:conditional}) 
in the limit of small grains.

The calculation of mixed second moments for pairs of species
that belong to different cliques,
such as
$\Nij{2}{3}$
involves probabilities such as
$P(N_2=1,N_3=1)$ for states in which
species that do not react with each other reside 
simultaneously on the grain.
It can be shown that
these probabilities do not satisfy the relation of
Eq. (\ref{eq:small_conditional}).
This result is consistent with the fact that
the multiplane method does not provide 
accurate results for this moment,
as already observed in 
Fig. \ref{fig5}(b). 

\begin{figure} 
\includegraphics[width=4.5in]{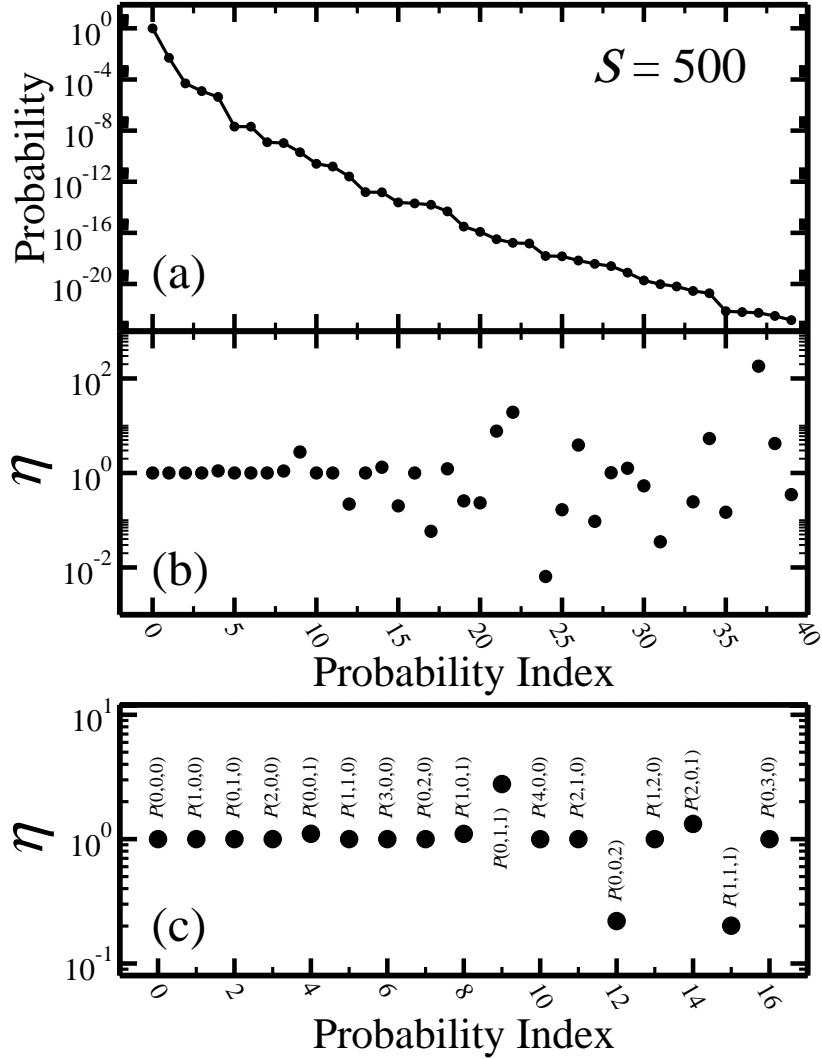}
\caption
{
(a)
The probabilities
$P(n_1,n_2,n_3)$,
obtained from the master equation,
arranged in descending order from the largest (left) to the
smallest (right)
for a small grain with 
$S = 500$
adsorption sites.
Here we show the forty largest probabilities;
(b)
The ratio parameter, 
$\eta$,
defined in 
Eq. (\ref{eq:eta}),
between the probabilities obtained from the 
multiplane equations and the corresponding 
probabilities obtained from the master equation.
The ratio is unity for the first few probabilities
and then it fluctuates for the the rest;
(c)
An enlargement of the first seventeen probabilities displayed
in (b).
Standing out are
$P(0,1,1)$
and 
$P(1,1,1)$,
for which the multiplane equations and the master equation differ.
These probabilities have no significant effect on the production rates and
population sizes, 
but are expressed when computing other moments.
}
\label{fig6}
\end{figure}

These results further support the conclusion that the multiplane method
is suitable for the calculation of moments confined to a single clique 
and is unsuitable for moments that combine different cliques.
To test this conclusion in detail 
we define the ratio

\begin{equation}
\eta (n_1,n_2,n_3) = {P_{\rm MP}(n_1,n_2,n_3) \over P(n_1,n_2,n_3)}, 
\label{eq:eta}
\end{equation}

\noindent
which is equal to 1 where the multiplane method is accurate and deviates
from 1 elsewhere. 
In Fig. 
\ref{fig6}(a) 
we display the forty highest probabilities,
$P(n_1,n_2,n_3)$, 
obtained from the master equation 
for the network of 
Fig. \ref{fig1}(b) 
in descending order.
The results are for a small grain of 
$S = 500$ 
adsorption sites,
for which the population sizes of adsorbed species are
exceedingly small.
The probabilities drop off very rapidly, 
implying that the first and second moments 
are indeed dominated
by the few highest probabilities.
In 
Fig. \ref{fig6}(b) 
we show the parameter
$\eta$, 
for the same set of probabilities.
It is confirmed that the multiplane method is 
valid only for the largest probabilities.
Beyond the first few entries, 
$\eta$ begins to fluctuate.
In 
Fig. \ref{fig6}(c) 
we show an enlarged plot of 
$\eta$, 
including the first 17 probabilities.
In this graph the probabilities are labeled.
It is found that for those probabilities associated with states 
in which only species from a single clique reside simultaneously
on a grain, the multiplane method is in excellent agreement
with the master equation. 
For states in which species from different cliques reside 
simultaneously on the grain, significant deviations are obtained.
The first significant deviation between the multiplane method
and the master equation is found for the probability
$P(0,1,1)$,
in agreement with the previous analysis.

The analysis above shows that 
the multiplane method is valid in the limit of small grains.
In this limit, the probability distribution is dominated
be a few high probabilities associated with small 
population sizes of the reactive species.
These dominant probabilities 
satisfy the approximation of 
Eq. (\ref{eq:conditional}).
Therefore, the  
population sizes and reaction rates obtained from
the multiplane method and the master equation
are in excellent agreement.

\subsection{The Limit of Large Grains}
\label{large_grains}

The applicability of the multiplane method in the 
limit of large grains is not surprising
because in this limit even the rate equations
provide accurate results.
As shown above,
the rate equations can be derrived from the
multiplane equations by removing the conditions from the
conditional averages. 
The accuracy of the rate equations shows that 
in the limit of large grains 
the correlations are negligible and 
the probability distribution
$P(n_1,n_2,n_3)$
can be factorized into a product of 
probabilities of single-species. 
Therefore, the multiplane method provides
accurate results for any desired moments
of the probability distribution.

\section{The Multiplane Equations for Complex Networks}
\label{complex}

\begin{figure} 
\includegraphics[width=5.0in]{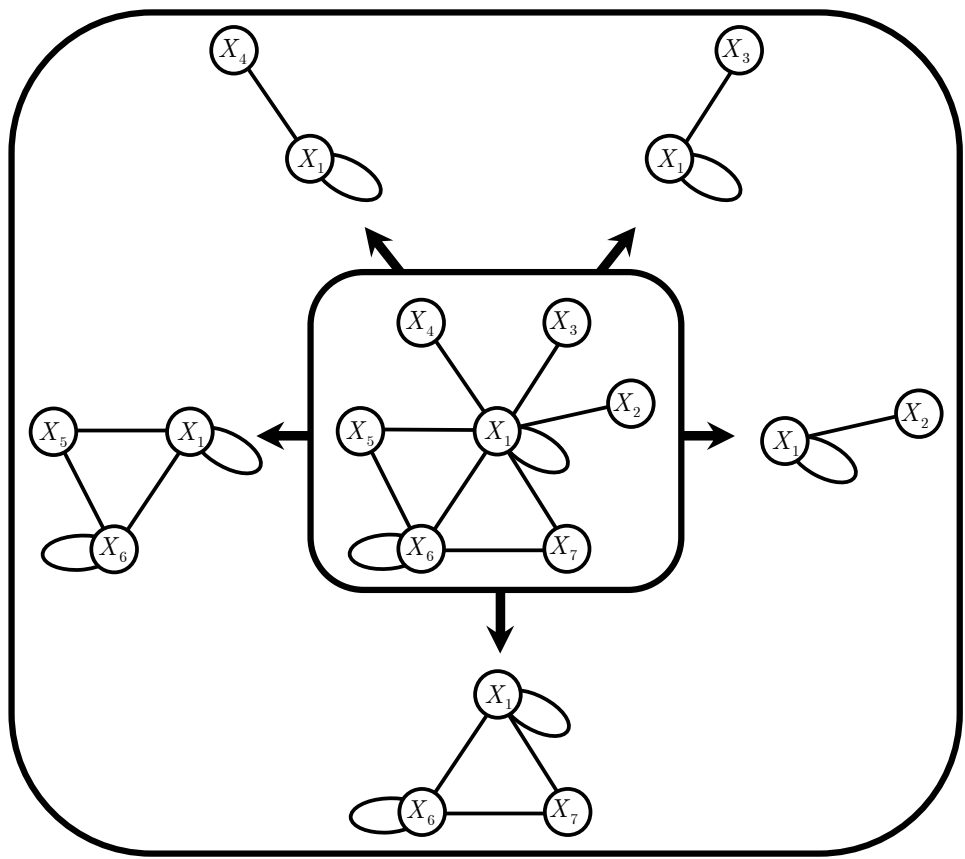}
\caption
{
A graph that represents
a complex reaction network which consists of seven
reactive species.
In the multiplane method, this network is broken into 
five cliques, and a lower dimensional master equation
is constructed for the marginal probability distribution
associated with each clique.
In interstellar chemistry, this is the network leading
to the formation of methanol on dust-grain surfaces.
}
\label{fig7}
\end{figure}

For sparse reaction networks with fluctuations, 
the multiplane method
was found to provide a dramatic reduction in the number of equations
compared to the master equation.
The method provides accurate results 
for the populations of reactive species 
and for the reaction rates. 
The method was presented for simple networks
which include only three species. 
However, the generalization to
more complex networks is straightforward.
Consider the network shown in 
Fig. \ref{fig7}. 
The probability distribution of the population sizes of the
reactive species in this network is 
$P(n_1, \dots, n_7)$.
To derive the multiplane equations 
one first needs to split the network into maximal cliques,
or maximal fully-connected subgraphs.
For the network of 
Fig. \ref{fig7} 
these cliques are:
$C_1:\{X_1,X_2\}$, 
$C_2:\{ X_1, X_3\}$, 
$C_3:\{ X_1, X_4\}$, 
$C_4:\{ X_1, X_5, X_6\}$
and 
$C_5:\{ X_1, X_6, X_7\}$.
The next step is to write down the master equation for
the marginal probability distribution associated with each clique. 
This can be done using either the top-down approach,
which is straightforward but tedious, 
or the bottom-up approach. 

In the top-down approach, 
the master equation for the marginal probability distribution
associated with a given clique is obtained by
tracing over all the species that do not belong to this clique.
This procedure is repeated for each one of the maximal cliques.

In the bottom-up approach, 
the master equation for 
the internal reactions in each
clique is constructed first.
Then, the coupling terms between cliques, 
which include the conditional
averages are added one by one.  
These terms account for reactions between species, such as $X_j$, 
which belong to the clique and species, such as $X_k$,  
which do not
belong to the clique.
In the master equation, the reaction between $X_j$
and $X_k$
is described by terms of the form
$n_j n_k P(\dots,n_j,n_k,\dots)$.
In the multiplane equation for the given clique,
$n_k$ is replaced by $\langle N_k \rangle_{n_j}$
and $P(\dots,n_j,n_k,\dots)$
is replaced by the marginal probability distribution 
for the clique.
For example, the resulting equation for the clique $C_1$
is

\begin{eqnarray}
\dot P(n_1,n_2) 
&=& F_1 [P(n_1 - 1,n_2) - P(n_1,n_2)] \nonumber \\
&+& F_2 [P(n_1,n_2 - 1) - P(n_1,n_2)] \nonumber \\
&+& W_1 [(n_1 + 1) P(n_1 + 1,n_2) - n_1 P(n_1,n_2)] \nonumber \\
&+& W_2 [(n_2 + 1) P(n_1,n_2 + 1) - n_2 P(n_1,n_2)] \nonumber \\
&+& (A_1 + A_2) 
[(n_1 + 1)(n_2 + 1)P(n_1 + 1,n_2 + 1) - n_1n_2 P(n_1,n_2)] \nonumber \\
&+& A_1 [(n_1 + 2)(n_1 + 1)P(n_1+2,n_2) - n_1(n_1 - 1)P(n_1,n_2)] \\
&+& \sum_{i=3}^{7}{(A_1 + A_i)[(n_1 + 1) 
\langle N_i \rangle _{n_1 + 1}P(n_1 + 1,n_2)
 - n_1 \langle N_i \rangle _{n_1}P(n_1,n_2)]}, \nonumber
\end{eqnarray}

\noindent
where 
Eq. (\ref{eq:conditional})
is used in order to justify the 
replacement of
$\langle N_i \rangle_{n_1,n_2}$ 
by
$\langle n_i \rangle_{n_1}$.
We find it instructive to carry out the procedure for the clique 
$C_5$
as well. 
In this clique, 
the species 
$X_1$
and 
$X_6$
are both correlated 
with 
$X_5$.
When tracing over $X_5$ one must maintain both correlations, 
giving rise to the conditional average
$\N{5}_{n_1,n_6}$.
The resulting equation takes the form:

\begin{eqnarray}
\dot P(n_1,n_6,n_7) 
&=& \sum_{i=1,6,7} F_i [P(\dots,n_i-1,\dots) - P(n_1,n_6,n_7)] \nonumber \\
&+& \sum_{i=1,6,7} 
W_i [(n_i + 1) P(\dots,n_i+1,\dots) - n_i P(n_1,n_6,n_7)] \\
&+& (A_1 + A_6) 
[(n_1 + 1)(n_6 + 1)P(n_1 + 1,n_6 + 1,n_7) - 
n_1n_6 P(n_1,n_6,n_7)] \nonumber \\
&+& (A_1 + A_7) 
[(n_1 + 1)(n_7 + 1)P(n_1 + 1,n_6,n_7 + 1) - 
n_1n_7 P(n_1,n_6,n_7)] \nonumber \\
&+& (A_6 + A_7) 
[(n_6 + 1)(n_7 + 1)P(n_1,n_6 + 1,n_7 + 1) - 
n_6n_7 P(n_1,n_6,n_7)] \nonumber \\
&+& \sum_{i=1,6} A_i 
[(n_i + 2)(n_i + 1)P(\dots,n_i+2,\dots) - 
n_i(n_i - 1)P(n_1,n_6,n_7)] \nonumber \\
&+& \sum_{i=2}^{4}{(A_1 + A_i)
[(n_1 + 1) \langle N_i \rangle _{n_1 + 1}P(n_1 + 1,n_6,n_7)
 - n_1 \langle N_i \rangle _{n_1}P(n_1,n_6,n_7)]} \nonumber \\
&+& (A_1 + A_5) 
[(n_1 + 1) \langle N_5 \rangle _{n_1 + 1, n_6} P(n_1 + 1,n_6,n_7) 
 -  n_1 \langle N_5 \rangle _{n_1, n_6} P(n_1,n_6,n_7)] \nonumber \\
&+& (A_5 + A_6) 
[(n_6 + 1) \langle N_5 \rangle _{n_1, n_6 + 1} P(n_1,n_6 + 1,n_7) 
 -  n_6 \langle N_5 \rangle _{n_1, n_6} P(n_1,n_6,n_7)]. \nonumber
\end{eqnarray}

\noindent
This network has been simulated using both 
the multiplane method and the complete master equation.
The results were found to be in excellent agreement
\cite{Lipshtat2004}.

\section{Applications to Physical and Biological Systems}
\label{applications}

Stochastic simulations are of great importance 
in a wide range of physical systems. 
Below we present two examples of current research areas 
in which the multiplane method is expected to be useful.

\subsection{Chemical Networks on Interstellar Grains}

The chemistry of interstellar clouds includes gas-phase reactions 
as well as grain-surface reactions
\cite{Hartquist1995,Tielens2005}.
Due to the microscopic size of the grains, 
and the low flux, the population
sizes of reactive species on the grains 
may be small and exhibit strong fluctuations.
Under these conditions rate equations are not suitable
for the simulation of grain-surface chemistry
\cite{Tielens1982,Charnley1997,Caselli1998,Shalabiea1998}.
To account correctly for the reaction rates, stochastic 
methods such as direct integration of the master equation 
\cite{Biham2001,Green2001,Biham2002}
or Monte Carlo simulations
\cite{Tielens1982,Charnley1997,Charnley2001}
are required.
The master equation is more suitable for grain chemistry
because it consists of differential equations, which can
be easily coupled to the rate equations of gas phase chemistry.
Furthermore, the master equation provides the probability 
distribution from which the reaction rates can be evaluated
directly, unlike Monte Carlo methods that require to accumulate 
large sets of data and to perform ensemble or temporal averages. 
For simple networks the master equation is efficient and provides
accurate results. 
However,
for complex networks, 
the master equation becomes infeasible. 
In this case, the multiplane method provides efficient
stochastic simulations.

Consider the network of 
Fig. \ref{fig7}.
Using the following substitutions
$X_1 \rightarrow \rm{H}; X_2 \rightarrow \rm{OH}; 
X_3 \rightarrow \rm{H}_3\rm{CO}; X_4 \rightarrow \rm{H}_2\rm{CO};
X_5 \rightarrow \rm{HCO}; X_6 \rightarrow \rm{O};
X_7 \rightarrow \rm{CO}$, 
this network coincides with the
reaction network that leads to 
methanol production on grains
in molecular clouds
\cite{Tielens1982,Charnley1997,Stantcheva2002,Stantcheva2003,Lipshtat2004}.
Current experimental effort is aimed at the evaluation of
the relevant rate constants for the surface 
diffusion, reaction and desorption
of the species involved in this network
\cite{Perets2005,Hidaka2007}.
These experiments include infra-red spectroscopy
as well as temperature-programmed desorption runs
using a mass spectrometer to detect the desorbed molecules.  
The resulting parameters, inserted into the multiplane
equations, will enable to evaluate the reaction rates in interstellar
environments and to compare the results with observations.
The multiplane method for this network provides 
a reduction in the number of equations from
about one million, using the master equation, 
to about one thousand equations.
For more complex networks the master equation 
becomes infeasible while the multiplane method
remains efficient. 

\subsection{Genetic Networks in Cells}

Another important field in which the 
multiplane method is expected to be useful
is the study of
genetic regulatory networks in cells.
These networks describe the transcription of mRNAs from
genes and their translation into proteins.
The regulation is performed at the transcriptional level
(by transcription factors that bind to the promoter site
of the regulated gene),
at the post-trancriptional level (e.g., by small non-coding RNAs)
and at the post-translational level (e.g., by protein-protein
interactions).
Analysis of these networks revealed modular structure.
In particular, modules or motifs which perform 
specific functions and repeatedly appear in different parts of the
network were identified
\cite{Milo2002,Milo2004,Yeger-Lotem2004}.
Common examples of such motifs are the 
autorepressor~\cite{Rosenfeld2002} 
and different versions of the
feed forward loop~\cite{Mangan2003}.
Other modules such as the
genetic switch~\cite{Ptashne1992} and 
the mixed-feedback loop~\cite{Francois2005}
also appear, but are not as common.

Genetic networks often exhibit strong fluctuations due to the
fact that some of the transcription factors appear in low
copy numbers.
Moreover, the transcriptional regulation is typically performed
by a small number of transcription factors which bind and unbind
to the promoter site at a fast rate.
This gives rise to strong fluctuations in the transcription rate
of the regulated gene.
Some modules, such as the autorepressor, the genetic switch
and the mixed-feedback loop include positive or negative
feedback mechanisms, which tend to enhance the role of
fluctuations.
In particular, the dynamics of the genetic switch system
was studied extensively using both deterministic and
stochastic methods
\cite{Gardner2000,Kepler2001,Cherry2000,Warren2004,Warren2005,Walczak2005,Lipshtat2006,Loinger2007}.
It was found that fluctuations play a crucial role in
this system.
While the analysis of small modules such as the genetic switch
can be done using the master equation,
it quickly becomes infeasible when larger networks are considered.
The implementation of the multiplane method in this
context is expected to provide a broader perspective on
the role of fluctuations in genetic networks.  
Recently, such fluctuations at the level of single cells
were measured experimentally using the  
green fluorescent protein 
\cite{Elowitz2002}. 
Such measurements are also expected to provide the effective
rate constants of the relevant processes in the cell.

\section{Summary and Discussion}
\label{summary}

We have shown that
the multiplane method provides efficient simulations 
of complex reaction networks with fluctuations,
for which the rate equations fail and the master
equation is infeasible.
The multiplane equations are obtained by 
breaking the network into maximal cliques
and writing down the set of
master equations for the marginal probability 
distributions of these cliques, 
with a suitable coupling between them.
For typical sparse networks, the method provides 
a dramatic reduction in the number of equations.
We found that the multiplane results
for the first and second moments, which account
for population sizes and reaction rates, respectively,
are in excellent agreement with
those of the complete master equation. 
It also accounts correctly for higher moments
which involve species from the same clique.
However, the method does not account correctly for second
and higher moments which include species from 
different cliques.

The numerical results are complemented by an asymptotic
analysis of the small and large grain limits.
A more rigorous analysis 
shows that the multiplane method is 
asymptotically exact  
in both limits
\cite{Barzel2007}.
It is performed in a more general setting, in which the
maximal cliques may be broken into smaller cliques.
In particular, one may break the entire network into 
cliques of two species each.
It is shown that even in this case, in the limits of
small and large grains, the method still
provides exact results for all the first moments
and for those second moments that involve species
in the same clique.

A related approach,
based on moment equations,
also provides efficient 
stochastic simulations of reaction networks
\cite{Barzel2007a}.
In this approach, one constructs differential equations
for the first and second moments of the probability 
distribution.
The number of equations is further reduced to 
one equation for each reactive species
(node) 
and one equation for each reaction (edge). 
Thus, for typical sparse networks the complexity of the stochastic 
simulation becomes comparable to that of the rate 
equations.
In applications such as interstellar chemistry, in which
the main objective is to calculate the reaction rates,
the moment equations are advantageous.
However, in systems such as genetic networks, in which
the probability distribution itself is of interest,
the multiplane method is required.

\clearpage

\clearpage

\end{document}